\documentclass[aps,twocolumn,showpacs]{revtex4}
\usepackage{epsf,graphics,graphicx,amsmath,amssymb,multirow,rotate}

\usepackage{color}
\definecolor{red}{rgb}{1,0,0}
\definecolor{green}{rgb}{0,1,0}
\definecolor{blue}{rgb}{0,0,1}

\begin{document}

\phantom{n}

\title{Delayed Choice Between Purely Classical States}

\date{\today}

\author{Jason A.C.~\surname{Gallas}}
\affiliation{Instituto de F\'\i sica,  
             Universidade Federal do Rio Grande do Sul, 
             91501-970 Porto Alegre, Brazil} 
\begin{abstract}
It is argued that Wheeler's insightful idea of delayed choice 
experiments  may be explored at a classical level,
arising naturally from number-theoretical conjugacies always
necessarily present in the equations of motion.
For simple and representative systems,
we illustrate how to cast the equations of motion in a form
encoding all classical states simultaneously through
a ``state  parameter''.
By suitably selecting the parameter one may project the system into any 
desired classical state.
\end{abstract}

\pacs{03.65.Ta, 
      03.65.Fd, 
      03.65.Wj, 
      42.50.-p  
      }
\keywords{Classical correlations, entanglement, delayed choice,
  quantum information}

\maketitle

The investigation of the relationship between quantum and classical states 
is experiencing  nowadays a renewed upsurge of interest driven 
by new pressing challenges related to technological needs 
arising from quantum 
engineering \cite{zurek03,peres04,az05,cbs05,palacios05,tonomura05,kus05}
and from quantum information processors which promise efficient 
solution to problems that seem
intractable using classical devices \cite{nmr,pra05}.
A recent issue of Nature  
is dedicated to these novelties where it is explained why,
``despite some remaining hurdles, 
the mind-bending and frankly weird world of quantum computers 
is surprisingly close'' \cite{nat06}.

A particularly fruitful bridge facilitating the understanding
of correlations between quantum and classical phenomena has been
Bohr's principle of complementarity \cite{bohr} stating that
quantum systems, or ``quantons'' \cite{ll88,e96} 
posses properties that are equally real but  mutually exclusive.
With this concept, the familiar wave-particle duality may be phrased 
more objectively as follows:
Depending on the experimental circumstance a quanton  behaves
approximately either as a classical particle or as a classical wave.
The standard way of exploiting quantons is by using interferometers
such as 
the traditional double-slit experiment of Young \cite{y1804},
specially as manifest in  the novel and  very ingenious recent 
implementations \cite{atto05,prosen05,jacquod05},
or the Mach-Zehnder setup.
In these frameworks,
the signature of wavelike behavior is the familiar interference
pattern, whereas the signature of particlelike behavior emerges
whenever one can discriminate along which way the interferometer
has been traversed.
Many interesting ``classical effects'' of interference have
been discussed in the literature \cite{gsw95,g95,qops}.
Inspired by a series of experiments proposed by Wheeler 
some years ago,
we wish to consider here another type of classical analogon 
of notions familiar from investigations of quantons.

In an insightful contribution Wheeler named, spelled out
and elucidated by seven examples the so-called ``delayed choice
experiments'', instigated by the following question 
concerning the screen with two slits \cite{w78,mw83}:
Can one choose whether the photon (or electron) shall have
come through both of the slits, or only one of them, after
it has {\it already\/} transversed this screen?
His motivation to ask this question was what he refers to
as a {\it pregnant sentence\/} of Bohr:
``...it... can make no difference, as regards observable
effects obtainable by a definite experimental arrangement,
whether our plans for constructing or handling the
instruments are fixed beforehand or whether we prefer to
postpone the completion of our planning until a later moment
when the particle is already on its way from one instrument
to another''  \cite{bohr49}.
Experimental confirmation followed soon \cite{hellmuth87,edz}.

Here we argue that the fruitful concept of delayed choice is no 
alien to classical physics and may be directly recognized and
retrieved  from the multivalued nature of the equations of 
motion of all discrete-time dynamical systems of 
algebraic origin \cite{nota1}.
In classical dynamical systems, delayed choice is tantamount
to orbital parameterization and arises thanks to number-theoretical 
properties  shared by the equations defining the infinite set 
of periodic orbits building the 
scaffolding of orbits known to underly both classical and quantum
dynamics\cite{gutz,bb,berry,keating}.

To illustrate how delayed choice works in classical systems
we first consider the paradigmatic logistic 
map \cite{verhulst}  $x \mapsto f(x)=a-x^2$.
For $a=4$ this system serves adequately for dynamics based 
computation, in particular to emulate logic gates,  
to encode numbers, and to perform specific arithmetic operations 
such as addition and multiplication on these numbers \cite{sinha}.
We consider a fully general situation however,
letting $a$ to be a free parameter.
In discrete-time dynamics, the period is the foremost discrete
(``quantized'') quantity \cite{jgpre}.
To illustrate concepts and methodology,
we first derive a pair of polynomials,  $P_3(x)$ and $S_3(\sigma)$,
encoding  simultaneously all possible orbits of period-$3$.
Period 3 is the quintessence period in dynamical systems.
For, existence of period 3 implies existence of all other periods 
because of the powerful theorem ``period three implies chaos'' \cite{jim}.

Period-$3$ orbits are obtained by 
composing the equation of motion $f(x)$ three times consecutively,
$f^{(3)}(x)\equiv f(f(f(x)))$, and isolating  lower-lying periods
by division  as usual \cite{jgpre}.
This yields the basic polynomial $H_3(x)$ with roots $x_i$
ruling all  possible period-3  motions
\begin{eqnarray}
 H_3(x) &=& \big({x-f^{(3)}(x)}\big) / \big({x-f(x)}\big)\cr
        &=&  x^6 -x^5 -(3a-1)x^4 +(2a-1)x^3 \cr
        &&  \qquad\quad 
            + (3a^2  -3a  +1)x^2  -(a-1)^2x  \cr
        &&  \qquad\quad 
        -a^3+2a^2-a+1.  \label{h3}
\end{eqnarray}
The degree of $H_3(x)$ reveals that we have to deal with 
two period-$3$ orbits, which for arbitrary $a$ are distinct.
So, generically we deal with two independent triplets of
numbers which are entangled among the roots of $H_3(x)$.
Although the coefficients of the sextic are relatively tame functions
of $a$, its roots are not.
From the infinity of  possible relative-sextic quantities,
phase-space dynamics knows precisely which adequate
combination of algebraically {\it conjugate\/} numbers 
to select in order to produce tame $H_3(x)$ polynomials.
The task now is to to find a means of efficiently disentangling
the pair of triples composing individual period-3 orbits,
i.e.~of decomposing the sextic into a pair of cubics.
This will be done now in two different ways.
The first is a procedure that is sometimes practical.
The other one works systematically.

\noindent
{\it Symmetric representation}:
The standard mathematical way of finding roots of polynomials
relies on computing  {\it discriminants\/} \cite{discri},
generalizing the well-known procedure to solve 
quadratic equations.
So,  the discriminant of $H_3(x)$ is 
$ \Delta_3 \equiv \left(4a-7\right)^3\left(16a^2-4a+7\right)^2$.
It contains a factor which is not a perfect-square,
thus a natural candidate for an extension field over 
which to attempt to factor the equation of motion. 
Indeed, introducing the radical 
$r\equiv(4a-7)^{1/2}$, over the relative-quadratic extension
${\mathbb Q}(r)$ of the rationals one finds a symmetric decomposition 
$H_3(x)=\psi_1(x)\psi_2(x)$, valid for arbitrary
values of $a$, where
\begin{eqnarray*}  
    \psi_1(x)    &=& x^3 -u x^2 -(a+v)x
                      + 1-av,   \label{psi1}\\
    \psi_2(x)    &=& x^3 -v x^2 -(a+u)x
                      + 1-au,   \label{psi2}
\end{eqnarray*}
and where $u\equiv(1+r)/2$, $v\equiv(1-r)/2$.
This symmetric pair of cubics expresses the orbits as conjugate factors 
of the radical $r$. Interchanging the branches of $r$
simply converts one orbit into the other.
These cubics already show the interesting property that we wish to
explore: the {\it individual formal representation\/} of either
$\psi_1(x)$ or $\psi_2(x)$ already contains in its structure 
information concerning all possible physical solutions.
Particular solutions emerge only when we fix the
branches of the radical $r$.
The above derivation is not helpful in general 
because the presence of non-quadratic factors in the discriminant 
does not automatically imply  factorizability. 
A method that works in general, for arbitrary periods, is the following.

\noindent
{\it Asymmetric representation}:
Independently of discriminants,
the individual orbits entangled in $H_3(x)$ may be sorted out 
systematically as follows.
Denote by $\xi$ any arbitrary root of $H_3(x)$.
To form a period-$3$ orbit,
such root must be obviously connected to two companion roots:
$\xi$, $f(\xi)$, $f^{(2)}(\xi)=f(f(\xi))$. 
These orbital points split $H_3(x)$ into cubics.
They  may be used to construct
the familiar trio of elementary symmetric functions
\begin{subequations}
\begin{eqnarray}
  \theta_1(\xi) &=&  \xi + f(\xi)  + f^{(2)}(\xi)   \label{soma}\\
  \theta_2(\xi) &=&  \xi f(\xi) + \xi f^{(2)}(\xi)
             + f(\xi)f^{(2)}(\xi)\\
  \theta_3(\xi) &=&  \xi f(\xi) f^{(2)}(\xi),       \label{produto}
\end{eqnarray}
\end{subequations}
which remain invariant under permutations of the orbital points.
The fact that $f(\xi)=a-\xi^2$, $f^{(2)}(\xi)=a-(a-\xi^2)^2$  and that 
$\xi$ is a root of $H_3(x)$ allows us to express any pair $\theta_m(\xi)$,
$\theta_n(\xi)$ in terms of the remaining member of the trio.
A fruitful choice is to express 
$\theta_3(\xi)$ and $\theta_2(\xi)$
in terms of the sum $\theta_1(\xi)$ of orbital points:
\begin{eqnarray*}
  \theta_1(\xi) &=&  -\xi^4 +(2a-1)\xi^2 + \xi +2a-a^2\\
  \theta_2(\xi) &=&  H_3(\xi) + \theta_1(\xi) + a +1
      =   \theta_1(\xi) +a+1\\
  \theta_3(\xi) &=&  (\xi+1)H_3(\xi) - a\;\theta_1(\xi) + a -1\cr
                &=& -a\;\theta_1(\xi) + a-1.
\end{eqnarray*}
These three symmetric functions define the key cubic
\begin{eqnarray}
P_3(x) &=& x^3 -\theta_1(\xi)\;x^2 + \big(\theta_1(\xi)+a+1\big)\;x\cr
      && \phantom{x^3}  + a\;\theta_1(\xi)-a+1,  \label{p3}
\end{eqnarray}
the equation of motion for the disentangled orbit.

Similarly,
calling $\eta$, $f(\eta)$ and $f^{(2)}(\eta)=f(f(\eta))$ the 
remaining triplet of roots of $H_3(x)$, one sees that
they obey the same functional relations above,
namely $\theta_1(\eta)$, $\theta_2(\eta)$ and $\theta_3(\eta)$.
As already mentioned, this triplet is in general 
distinct from $\theta_1(\xi)$, $\theta_2(\xi)$ and $\theta_3(\xi)$,
since they define different orbits.

Denoting indistinctly by $x_1$, $x_2$, $\dots, x_6$ the roots of $H_3(x)$,
the sum and product of $\theta_1(\xi)$ and $\theta_1(\eta)$ 
are then
\begin{eqnarray*}
  \theta_1(\xi) + \theta_1(\eta) &=& \sum x_j = 1,\\
  \theta_1(\xi)\;\theta_1(\eta)
     &=& \sum_{j<k} x_jx_k + \theta_2(\xi)  +\theta_2(\eta)\cr
   &&\cr
     &=& -(3a+1) + \theta_1(\xi) + \theta_1(\eta) +  2(a+1)\cr
   & =& 2-a.
\end{eqnarray*}
These two quantities define a quadratic, say
    $ w^2 -w +2-a =0$,
with roots $ w = (1 \pm \sqrt{4a-7})/2$.
They are the numbers $\theta_1(\xi)$ and $\theta_1(\eta)$ 
needed in Eq.~(\ref{p3}) to obtain the pair of period-$3$ orbits. 
Instead of $w$ we introduce a more convenient parameter $\sigma$ 
through the transformation
$4a-7 = (2\sigma-1)^2$ or, equivalently, $S_3(\sigma)=0$ where
\begin{equation}
       S_3(\sigma) =  \sigma^2 -\sigma +2 - a,          \label{hhh}
\end{equation}
a polynomial that coincides with the above polynomial in $w$.
The solutions of $w^2 -w -a+2 =0$ may be also written as
        $w = ( 1 \pm 2\sigma \mp 1)/2$,
yielding the final answers in a very convenient form:
\begin{equation}
      \theta_1(\xi)  = \sigma \quad\qquad\hbox{and}\qquad\quad  
      \theta_1(\eta) = 1-\sigma,  \label{s1}
\end{equation}
or $\theta_1(\xi)  = 1-\sigma$ and  $\theta_1(\eta) =\sigma$. 
Recalling Eq.~(\ref{soma}) one sees that, for each periodic orbit,
the convenient parameter  $\sigma$ is simply the
{\it sum\/} of its orbital points.
Using the constraint $S_3(\sigma)=0$ we may eliminate $a$
from Eq.~(\ref{p3}) and obtain an equation whose unknown 
coefficient is either $\theta_1(\xi)$ or $\theta_1(\eta)$,
depending which orbit we want to consider.
For the choice  in Eq.~(\ref{s1}) we get
\begin{eqnarray*}
 \varphi_1(x)\equiv \varphi_1(x;\sigma) 
    &=& x^3 -\sigma\,x^2 -(\sigma^2-2\sigma+3)\,x\cr
    &&  \qquad  + \sigma^3-2\sigma^2+3\sigma-1,
                               \label{c1}\\
 \varphi_2(x)\equiv \varphi_2(x;\sigma) 
    &=& x^3 -(1-\sigma)\,x^2 -(\sigma^2+2)\,x\cr
    && \qquad  
    -\sigma^3+\sigma^2-2\sigma+1,
            \label{c2}
\end{eqnarray*}
yielding the $\sigma$-sextic $Q_3(x) = \varphi_1(x)\varphi_2(x)$, namely
\begin{eqnarray*}
  Q_3(x) &=& x^6 -x^5 + (-3\sigma^2+3\sigma-5)\,x^4\cr
     && +(2\sigma^2-2\sigma+3)\,x^3\cr
     && +( 3\sigma^4-6\sigma^3+12\sigma^2-9\sigma+7)\,x^2\cr
     && - (\sigma^4-2\sigma^3+3\sigma^2-2\sigma+1)\,x\cr
     && -\sigma^6+3\sigma^5 -7\sigma^4+9\sigma^3
    -9\sigma^2 + 5\sigma-1. 
\end{eqnarray*}
The cubics $\varphi_1(x)$, $\varphi_2(x)$ look very different
from $\psi_1(x)$, $\psi_2(x)$ although both pairs  
represent the same physics, i.e.~the same set of orbits.
By eliminating $\sigma$ between $S_3(\sigma)$
and either $\varphi_1(x)$ or
$\varphi_2(x)$ we get back the original polynomial $H_3(x)$
of Eq.~(\ref{h3}). 
Identical result is obtained eliminating $r$ between $r^2=4a-7$ and  
either  $\psi_1(x)$  or $\psi_2(x)$.
Comparing coefficients of equal powers in the sextics
$H_3(x)$ and $Q_3(x)$ 
one recognizes that all coefficients are 
interconnected by the constraint $S_3(\sigma)$.

The discriminants of the $\varphi_1(x)$ and $\varphi_2(x)$  
are
\begin{equation*}
   \Delta \varphi_1 = (4\sigma^2-6\sigma+9)^2,  \qquad
   \Delta \varphi_2 = (4\sigma^2-2\sigma+7)^2.
\end{equation*}
Now, recall  that ``any third-degree
polynomial $p(t) \in {\mathbb Q}(t)$ which is
irreducible over the rationals ${\mathbb Q}$ will have a {\it cyclic\/}
Galois group if and only if the discriminant of $p(t)$
is a square over $\mathbb Q$'' \cite{stewart}.
Thus, $\Delta \varphi_1$ and  $\Delta \varphi_2$ manifest  clearly
the advantage of $\sigma$-parameterization:
It produces at once orbital equations with cyclic Galois
group in a  number-field of degree coinciding with the period 
of the orbits, i.e.~the smallest number-field possible, 
yielding separated rather than entangled factors.
This is not the case if we compute discriminants for 
$\psi_1(x)$ and  $\psi_2(x)$.
Although $\varphi_1(x)$ and $\varphi_2(x)$ are distinct functions,
their discriminants with respect to $\sigma$ are identical.

After this {\it excursus\/} emphasizing strength 
and generality of the method, let us consider 
what is  encoded into $\varphi_1(x)$ and $\varphi_2(x)$.
As the constraint $S_3(\sigma)=0$ shows, 
two different values of $\sigma$ lead to the same value of $a$. 
For instance, by taking either $\sigma=0$ or $\sigma=1$ 
we reach $a=2$,  the ``partition generating limit'' with many valuable  
properties \cite{gutz},
the limit where one may emulate logic gates, encode numbers, 
perform specific arithmetic operations on these numbers \cite{sinha}, 
and more \cite{jgpre}.
For $\sigma=0, 1$ we dispose of two independent
microscopic $\sigma$-representations for each macroscopic state,
namely
\begin{eqnarray*}
           \Phi(x) &=& \varphi_1(x;0)
                    = \varphi_2(x;1) = x^3 -3x-1,\\
         \smallskip
\overline{\Phi}(x) &=& \varphi_1(x;1)
                    = \varphi_2(x;0) = x^3 -x^2 -2x +1,
\end{eqnarray*}
where the overline is used to indicate that, in spite of their rather 
different functional forms, both functions are dynamically conjugated.
These four expressions show that by permuting the
values of $\sigma$ we effectively {\it interchange\/}
orbits, independently of the choice for  $\varphi_\ell(x)$.
Macroscopically in phase-space we deal with 
$\Phi(x)$ and $\overline{\Phi}(x)$. 
But microscopically the description may be done equally well
using either $\varphi_1(x,\sigma)$ or $\varphi_2(x,\sigma)$.
This degeneracy is not normally seen in phase-space \cite{jgnew}.

Note that knowledge of just a single state, 
here $\varphi_1(x,\sigma)$ or $\varphi_2(x,\sigma)$,
is enough to grant access to all physical states 
because the results obtained for one of them follow automatically
for all {\it conjugate family\/} when we change the value of $\sigma$.
For higher periods, conjugate families normaly contain hundreds of
states.
This is quite a lot because period-three implies chaos \cite{jim}.
Thus, $\sigma$-the encoding 
stores conveniently all information concerning period-$k$ 
dynamics for arbitrary $k$.
It is a generic property of algebraic equations, not a 
peculiarity of the illustrative example considered.
By suitably
selecting $\sigma$ one may switch from one orbit to another, 
performing a ``delayed choice'' of the ordering (labeling).
By iterating polynomial automorphisms rather than orbital points one
my even bypass the need for finding orbits in phase-space.
In other words,  orbits are automorphically correlated  \cite{jgnew}.
 It is as if we were dealing with a multilevel ``atom'' in which the
 states could be defined and redefined  by selecting
 the appropriate $\sigma$. 

Do such parametric encodings exist also in more complicated
multidimensional dynamical systems?
Yes: the algebraic properties explored here are
generic for dynamical systems of algebraic origin \cite{nota1}.
Incidentally, one-dimensional systems do not represent any 
restriction because multidimensional systems may be always 
reduced to one-dimensional equivalents \cite{eg2002}.
For example, for the H\'enon map $(x,y)\mapsto (a-x^2+by,x)$,
the prototypical multidimensional system which among other things
describes very well the parameter space of class B lasers,
CO$_2$ lasers in particular \cite{bgg,eg06}, 
the generic cubic orbit encoding all period-$3$ solutions and valid
for arbitrary values of the parameters $a$ and $b$ is
\begin{eqnarray*}
  {\mathcal P}_3^H(x) 
   &=& x^3 -\sigma x^2 
      -\big[\sigma^2-2(1-b)\sigma + 3(1+b+b^2)\big]x\cr
   &&  + \sigma^3-2(1-b)\sigma^2 
       + (3+b+3b^2)\sigma-1+b^3,
\end{eqnarray*}
where $\sigma$ is now any root of the quadratic
\begin{equation*}
  {\mathcal S}_3^H(\sigma) = \sigma^2 - (1-b)\sigma + 2(1+b+b^2) - a.
\end{equation*}
In the fully dissipative $b=0$ limit
these equations correctly reproduce all results above.
Parameterized equations covering the Hamiltonian limit ($b=-1$) 
and valid for all periods up to 22 are studied elsewhere \cite{egPLA}.
Thus,  one clearly sees that
adding more parameters and/or extra dimensions only 
alter coefficients, not substance.

By adapting a concept developed for understanding 
quantum measurements we obtained a unified picture of
what happens at the micro and macroscopic level of 
discrete-time classical dynamical systems.
This new perspective is of course expected to apply
equally well to more general situations, 
not only to algebraic systems. 
Although mathematical difficulties in deriving closed-form results 
for more intricate equations of motion greatly increase
in this very general setup, 
no essential hindrances are anticipated to exist.

The author thanks CNPq, Brazil, 
for a Senior Research Fellowship.  

\end{document}